\font\mybb=msbm10 at 12pt
\def\bbxx#1{\hbox{\mybb#1}}
\def\Z {\bbxx{Z}}
\def\R {\bbxx{R}}

\def \aa {\alpha}
\def \bb {\beta}
\def \gg {\gamma}
\def \dd {\delta}
\def \ee {\epsilon}

\def \mm {\mu}

 \def \ggg {\Gamma}

\def \2 {{1 \over 2}}
\def \3 {{1 \over 3}}
\def \4 {{1 \over 4}}
\def \5 {{1 \over 5}}
\def \6 {{1 \over 6}}
\def \7 {{1 \over 7}}
\def \8 {{1 \over 8}}
\def \9 {{1 \over 9}}
\def \0 { \infty}

\def\++ {{(+)}}
\def \- {{(-)}}
\def\+-{{(\pm)}}

\def \pa {\partial}

\documentstyle[prd,aps,preprint,tighten,floats]{revtex}
\begin{document}
\draft
\preprint{ UPR-760-T, QMW-97-27, hep-th/9709033}
\date{September 1997}
\title  {Wrapped Branes and Supersymmetry}

\author{Mirjam Cveti\v c$^{1}$ and Christopher M. Hull$^2$
}
\address{$^1$\ Department of Physics and Astronomy \\
          University of Pennsylvania, Philadelphia PA 19104-6396,\\and\\
          $^2$  Physics Department, Queen Mary and Westfield College \\
 Mile End Road, London E1 4NS, U.K.\\  }\maketitle
\begin{abstract}
{Configurations of two or more branes wrapping different homology cycles of
space-time are
considered and the amount of supersymmetry preserved is analysed, generalising
the analysis of
multiple branes in flat space.
For $K3$ compactifications, these give the Type II or M theory origin of
certain supersymmetric
four-dimensional  heterotic string solutions that fit into
 spin-$3/2$ multiplets and which  become massless at certain points in moduli
space.
The interpretation of these BPS states and the possibility of supersymmetry
enhancement are
discussed.
 }
\end{abstract}
\newpage
\section{Introduction}

Branes have played a crucial role in unraveling the non-perturbative structure
of string theory
and M theory~\cite{HT}.
In particular,  $p$-branes which break half of  the supersymmetry play a key
role.
They can be combined to form bound states and there are
configurations in which they intersect or overlap
either perpendicularly~\cite{PT,Gaunt} or at   angles~\cite{BDL,GGPT,BC}
(For a
review see, e.g.,~\cite{Tseyrev,Gauntrev}.).
Both types of configurations, and in particular
the amount of supersymmetry preserved by  them, have been the subject of much
study
~\cite{Douglas,Polch,Green,Gaunt,GGPT,CC,Gauntrev,Pope}.

Branes can also  wrap homology cycles of a compactifying space to give BPS
states  in
the compactified
theory that become massless when the cycle degenerates. When the BPS states
that become massless
fit into a vector multiplet, this results in an enhancement of the gauge
symmetry~\cite{HUSC,Witten,HTE,BSV,BBO,strom,BLS}.

The purpose of this paper is two-fold. First we wish to extend the study of
multiple brane
configurations and supersymmetry to investigate systems in which various branes
wrap various
{\it different }{ homology cycles}. For toroidal backgrounds, the properties of
such
configurations
follow immediately from the corresponding properties of intersecting or
overlapping infinite branes
in flat space, but for branes wrapping different cycles of $K3$ or a
Calabi-Yau manifold, the
situation is more complicated. In particular, it requires a suitable
``alignment'' between the
calibrations of the various cycles.
This is important for the  study of symmetry enhancement with gauge groups
of rank of at
least two, resulting from two or more cycles degenerating simultaneously.
We shall demonstrate  the existence of  {  supersymmetric, composite}
  configurations, consisting of  two  branes    wrapping
round  two different  (shrinking) cycles of $K3$, with each brane  separately
breaking half of the
original
supersymmetry.

Second, we wish to investigate whether supersymmetry can be enhanced in the
way that gauge
symmetry can, with  BPS states fitting into multiplets containing spin-$3/2$
becoming massless
at special points in moduli space. In~\cite{HTE}, such supersymmetry
enhancement of,
e.g.,   $N=4$
supersymmetry in four dimensions ($D=4$),
was shown to be
impossible without running into problems of higher spin or infinite numbers of
massless fields.
However, at enhanced symmetry points of the heterotic string with $N=4$
supersymmetry in $D=4$, e.g.,
the vector supermultiplets, containing  the ``W-bosons'' that
become massless are
accompanied by an infinite
number of magnetic
and dyonic S-duality partners that also become massless~\cite{HTE}, so that the
result
cannot be described
by a field theory; in such exotic situations,
the possibility of supersymmetry
enhancement
deserves re-examination. In~\cite{CYII,ChC}, it was shown that there  exist
certain
supersymmetric classical  solutions  of
$N=4$ supergravity coupled to 22 vector multiplets in $D=4$ (effective theory
of
heterotic string compactified on $T^6$)
 that fit into
multiplets containing
spin-$3/2$
and whose  BPS mass formula  implies that they should become massless at
special points in
moduli space. These are configurations (and their S-duals) whose  electric {
and}
magnetic charge vectors each have length-squared $-2$. If the non-perturbative
heterotic
string indeed contains (quantum)  BPS
states   with the
charges carried by these configurations
then these states fit into
multiplets containing a
gravitino that becomes massless at certain points in moduli space. This  would
then result in
  an enhancement of the supersymmetry.
The points in moduli space at which this could happen~\cite{ChC}
 involve gauge groups
of
rank at least two.
 Examples of such massless configurations  are
associated  with branes wrapping {\it  two or more cycles} of $K3$($\times
T^2$)
which degenerate
simultaneously, so that
in this case the analysis involves studying various branes wrapping various
cycles, as discussed
above.
As is clear from the Type II description, each such $D=4$ supersymmetric
classical
configuration is the coincidence limit of a two-centre
 solution and corresponds to
the limit of a two-particle state in which two BPS states
(with no mutual force) approach one
another. If a bound state is formed, it must fit into a spin-$3/2$
 multiplet which becomes massless
at special points in the moduli space, which would result
in supersymmetry enhancement.
If no bound state is formed, there is no supersymmetry enhancement.

The equivalence~\cite{HT}  of Type IIA string theory compactified on $K3$,
and heterotic
string compactified on $T^4$
implies the existence of special points in the $K3$ moduli space at which the
gauge symmetry is
enhanced~\cite{Witten,HTE,BSV,BBO}. Indeed, it follows from supersymmetry that the
mass of
certain BPS states, preserving half of the supersymmetry,  tends to
zero at these special points~\cite{HUSC,HTE} to give the extra massless vector
multiplets. At these
points the area of certain homology two-cycles shrinks to zero to give an
orbifold limit of $K3$ and
the BPS states in $D=6$ arising from 2-branes wrapped round the
shrinking two-cycles become
massless ~\cite{Witten,HUSC,HTE}.

The behaviour of the theory as a particular  set
of two-cycles shrinks
can be studied by looking at the theory on the   space $\R^{(D-5,1)} \times
N_4$
in which the $K3$ is
replaced by the appropriate
asymptotically locally Euclidean (ALE) multi-instanton solution $N_4$ (e.g.,
 for
$SU(N)$ gauge
symmetry, this is a
 multi-Eguchi-Hanson space).
This  space gives a good
approximation  when
the radius of the  two-sphere is small compared to the size of the $K3$.
 The
enhanced gauge symmetry
is then associated with parallel branes becoming coincident, so that homology
two-spheres shrink to
zero size and the  membranes wrapped round these cycles
 give rise to massless states on the brane, whose   world-volume theory
thus becomes a
non-Abelian gauge theory.   This result is straightforward to check using
duality; a
T-duality relates this
solution to a multi-centre solitonic 5-brane solution of Type IIB theory (in
$\R^4/ \Z_s$) and $SL(2,\Z)$ duality relates this to a     configuration of
parallel 5-D-branes, and
there is symmetry enhancement as these become coincident due to strings joining
the 5-D-branes
becoming massless~\cite{BSV}.

Kaluza-Klein (KK)
monopole space-times are again of the form
 $\R^{(D-5,1)} \times N_4$, but with
$N_4$ now a   $D=4$ asymptotically locally flat (ALF)  instanton solution.
In~\cite{Hull}, it was argued that
KK
monopoles can be regarded as branes (6-branes in M theory or 5-branes
in $D=10$ string theory) and that
there is a similar symmetry enhancement
as KK monopole branes become coincident.
Again, in the coincident limit, certain two-cycles shrink to zero area and
branes
wrapping the
cycles become null.

In this paper  we shall generalise this  to consider
various branes wrapping various
{\it different cycles} and analyse the surviving supersymmetry for such BPS
configurations.
We shall concentrate on ALE or ALF instantons, and the implications for $K3$
compactifications,
although the generalisation to other spaces should be straightforward. In
Section II we summarize the
properties of the  multi-instanton metric
$N_4$, the role it plays in repairing orbifold singularities of $K3$ and the
conditions on
cycles to be supersymmetric.  In Section III we  review the   supersymmetry
constraints for  wrapped brane  configurations.  In Section IV we  address
multiple wrapped branes and the number of   preserved supersymmetries. In
Section V
we relate the   results in the Type II  string picture to those in the  dual
heterotic string picture.  In Section VI we
discuss the possibility of   enhanced supersymmetry.

\section{Calibrations and Gravitational Instantons}

We shall restrict ourselves to supersymmetric $p$-branes which break  half the
supersymmetry
when embedded as an
infinite brane in flat space, or wrapped round a $p$-torus. The main example
will be the D-branes
of Type II string theory. If a $p$-brane is wrapped round an
$n$-cycle ($n\le p$) in a space-time $N$, then   the amount of supersymmetry
preserved depends on
the cycle chosen. (The classification of supersymmetric cycles  on
Calabli-Yau spaces has been
studied by several groups in Refs. \cite{BSV,BBO}.). A brane wrapped round a
   {  supersymmetric cycle}  preserves some fraction of the supersymmetry and
the cycle  is  that of minimal
$n$-volume in a given homology class. The volume form of the supersymmetric
cycle is given
by the  pull-back of a certain closed form $\omega$ in $N$, which is a {\it
calibration } of the
cycle.

For a Calabi-Yau $n$-fold, there are $n$-dimensional supersymmetric cycles
that are  calibrated by
$Re(\Omega)$, where
$\Omega$ is the holomorphic $(n,0)$-form, and these cycles are said to be
special
Lagrangian.
There are also {  holomorphic} $2m$-cycles which are calibrated by ${1 \over
m!} J^m$,
where $J$ is the K\"ahler form. A $p$-brane wrapped round a holomorphic cycle
or
a   special
Lagrangian cycle will preserve half the supersymmetry.
For    four-dimensional hyper-K\"ahler manifolds $M$, the supersymmetric
two-cycles
are
 the holomorphic
two-cycles, i.e. those for which the volume form is proportional to the
pull-back
of one of  complex
structures of $M$.
\vskip 3mm
\noindent{\it ALE and ALF Spaces}
\vskip 3mm
The Gibbons-Hawking gravitational multi-instanton metric~\cite{GH} is
\begin{equation}
ds^2=V(d y + A_i d x^ i)^2 +V^{-1}\dd_{ij}d x^i
d x^j,
\label{GH}
\end{equation}
where $i=1,2,3$ and
\begin{equation}
 V^{-1}=\epsilon
+\sum _{r=1}^s {2n  \over \vert x^i-x_r^i\vert },
\qquad
\nabla_ i A_j-\nabla_jA_i=\ee
_{ijk}\nabla^k V^{-1}.
\label{}
\end{equation}
If  $\epsilon=1$, this is the multi-Taub-NUT space   with asymptotically
locally  flat (ALF) boundary
conditions, while if $\epsilon=0$, then $n$ can be scaled to $1$ and the space
is a multi-centre generalisation of the
Eguchi-Hanson metric with asymptotically locally  Euclidean (ALE) boundary
conditions. The case $\epsilon=0, s=1$ is flat space, while
$\epsilon=0, s=2$ gives the Eguchi-Hanson instanton.
The space  $\R^{(D-5,1)} \times N_4$ where $N_4$ is an  ALF or ALE instanton
is a solution of M theory $(D=11)$ or of string
theory ($D=10$), respectively, as the gravitational instanton is hyper-K\"
ahler~\cite{AGH}. It can be thought of as a set of $D-5$ branes which will be
referred
to as G-branes, following~\cite{Hull}.

The multi-Eguchi-Hanson space behaves as $\R^4/ \Z_s$ at large distances, and
can be used to resolve
an $A_{s-1}$ singularity of an orbifold limit of $K3$. For example, one
orbifold
limit of $K3$ is
$T^4/\Z_2$ and each of the 16 orbifold singularities can be repaired by gluing
in an Eguchi-Hanson
metric. In the limit $\vert x_1-x_2\vert  \to 0$, the Eguchi-Hanson space
becomes $\R^4/ \Z_2$ with
an orbifold singularity at the origin.  While the multi-Taub-NUT solution can
be thought of as a
solution with a number of parallel $D-5$ G-branes, the multi-Eguchi-Hanson
space can be viewed as a
number of parallel $D-5$ G-branes
 in  a transverse space which is an orbifold; as a result, not all the branes
are independent as
some are  ``mirror images'' of others. There is one instanton corresponding to
each pair
$\{x_r,x_s\}$.

As the $y$ coordinate is periodic, each line segment in the $\R^3$
parameterised by $x^i$ is
associated with a cylinder in
$N_4$, unless the line segment passes through one of the points $x_r^i$ at
which the size of the
$y$-circle shrinks to zero. In particular, a line segment joining $x_r^i$ and
 $x_s^i$
corresponds to a two-sphere and the set of all such two-spheres corresponding
to
all pairs $\{x_r,x_s\}$
forms a basis for the second homology.
As one approaches a point in moduli space at which $\vert
x_r-x_s\vert
\to 0$, the area of the corresponding two-sphere shrinks to zero.
The minimal two-cycle (the one with least area) in the homology class,
corresponding to the pair
$\{x_r,x_s\}$, is given by the orbit of the straight line in $\R^3$ joining
$\{x_r,x_s\}$ under the
$U(1)$  generated by the Killing vector $\pa/\pa y$.
For this to be a supersymmetric cycle, however, it is necessary that this
cycle be holomorphic.

These hyper-K\"ahler spaces have three complex structures $(J^a)_i {}^j$
($a=1,2,3$)
and three K\"ahler forms $J^a={1\over 2}(J^a)_{i  j}dx^i _\wedge dx^j$
given by~\cite{comst}
\begin{equation}
J^i=(dy+A_jdx^j)_\wedge dx^i - {1\over 2} V\ee^{ijk}dx^j_\wedge dx^k,
\label{csI}
\end{equation}
where the index $a$ has been identified with the spatial index $i$.
Consider the cycle corresponding to the pair $\{x_r,x_s\}$. Defining the
normalised direction
\begin{equation}
n^i_{rs} ={x^i_r-x^i_s
\over \vert x^i_r-x^i_s\vert },
\label{nordir}
\end{equation}
the minimal cycle $\{x_r,x_s\}$ is holomorphic with respect to the complex
structure~\cite{comst}
\begin{equation}
J_{rs}=n^i_{rs}J_i.
\label{csII}
\end{equation}

In particular, the  two-cycle  $\{x_r,x_s\}$ and the
two-cycle  $\{x_t,x_u\}$ will only be holomorphic with respect to the {\it
same}
complex structure (\ref{csII})
if the corresponding line segments are parallel, i.e. $n^i_{rs} =\pm n^i_{tu}$.

\section{Wrapped Branes and Supersymmetry}

Consider the space-time $\R^{(3,1)}\times N_4\times T^2$ where $N_4$ is an ALE
or ALF
instanton with
metric (\ref{GH}), regarded as a solution of Type IIA or Type IIB string
theory.
 We choose coordinates $X^M$ with $M=0,\cdots,9$, which include  the
coordinates
on
$N_4$, $X^\mu$, $\mm=1\cdots,4$, with $X^4=y$
and $X^i=x^i$ ($i=1,2,3$). The coordinates on the two-torus $T^2$ are $X^5,X^6$
and the coordinates
on the Minkowski space $\R^{(3,1)}$ are $X^0, X^a$ with $a=7,8,9$. The fact
that
$N_4$ is self-dual
restricts the supersymmetry parameters $\ee$ to those satisfying
\begin{equation}
\ggg _{N_4} \ee=\ee,
\label{scI}
\end{equation}
where $\ggg_{N_4}$ is the chirality projector on $N_4$:
$\ggg _{N_4} \equiv\ggg _1 \ggg _2\ggg _3\ggg _4$ 
and $\ee$ is a 32-component Majorana spinor in the type IIA case, and is
a   doublet of Majorana-Weyl spinors $\ee^i$ ($1=1,2$) of the same 
chirality   for Type IIB
 string theory.

Consider a minimal two-sphere $S_{rs}$ corresponding to the pair of points
 $\{x_r,x_s\}$. Coordinates on $S_{rs}$ can be taken to be the coordinate
$n_{rs}^ix^i$ along the line segment joining $\{x_r,x_s\}$, together with $y$. For the type IIA theory,
 a 2-brane wrapped round $S_{rs}$ will be preserved under those
supersymmetries whose
parameter satisfies
\begin{equation}
\ggg _{(2)}\ee  = \ee ,
\label{scII}
\end{equation}
where
\begin{equation}
\ggg_{(2)} = {1\over 36}\ee ^{\aa \bb\gg } \pa_\aa X^M \pa _\bb X^N \pa _\gg
X^C
\ggg_{MNP},
\label{Gammatwo}
\end{equation}
and $\aa=0,1,2$ labels the 2-brane world-volume indices.
Thus for the 2-brane,
\begin{equation}
\ggg ^0n_{rs}^i\ggg ^i\ggg^4  \ee  = \ee 
\label{scIII}.
\end{equation}
This condition, together with (\ref{scI}), implies that
the wrapped 2-brane preserves $1/4$ of the original 32 Type II supersymmetries,
or $1/2$ of the 16  supersymmetries for
Type II on $\R^{(3,1)}\times N_4\times T^2$.

For a 4-brane wrapped round
$S_{rs}\times T^2$, the condition (\ref{scII}) becomes
\begin{equation}
\ggg ^0n_{rs}^i\ggg ^i\ggg^4 \ggg^5\ggg^6 \ggg^{11}\ee = \ee.
\label{scIV}
\end{equation}
For the type IIB string,
a 3-brane wrapping
$S_{rs}\times S^1$ ($S^1$ is along, say,  $X^5$)  the
condition (\ref{scII}) becomes
\begin{equation}
\ggg ^0n_{rs}^i\ggg ^i\ggg^4 \ggg^5  \ee^1 = \ee^2.
\label{scV}
\end{equation}
As before, the above  two configurations also preserve $1/4$ of the
supersymmetry of
the original Type II superstring, or  $1/2$ of of the supersymmetry for Type II
on
$\R^{(3,1)}\times N_4\times T^2$.
It is straightforward   to generalise the supersymmetry constraints to other
branes.

For the Type IIA string, the dimensional reduction to $D=4$ of the three-form
$C_{MNP}$ via the Ansatz $C=\sum _{rs}A^{rs}_{\ \wedge}  X_{rs}$
gives a  vector field
 $A^{rs}$   corresponding to
each two-cycle $S_{rs}$, where $X_{rs}$  is
the harmonic two-form dual to the two-cycle $S_{rs}$.
 A two-brane wrapped
round
$S_{rs}$ gives  a BPS state in $D=4$
which
preserves $1/4$ of the original 32 Type IIA supersymmetries and
is   electrically charged with respect to $A^{rs}$, while the four-brane
wrapped round
$S_{rs}\times T^2$ gives a 0-brane that is magnetically charged with respect to
$A^{rs}$.
The mass of these BPS states depends on the area of $S_{rs}$ (and the value of
the two-form gauge
field $B_{MN}$ on $N_4$~\cite{Aspin}) and will vanish at a particular point in
moduli space at
which the area of  $S_{rs}$
vanishes. The electrically charged states correspond to the ``W-boson''
supermultiplets
that
become massless  as
$U(1)$ symmetry associated with $A^{rs}$ is enhanced to $SU(2)$.
At the same time the magnetic monopoles from the wrapped 4-branes also
become massless~\cite{HTE}.
Indeed, a whole $SL(2,\Z)$ multiplet of dyons become massless
 at the same point of
moduli space~\cite{HTE}. These arise
as follows. In flat space, a 2-brane with charge $q$   lying
inside a 4-brane of
charge $p$
  form a   bound state that is a BPS state breaking half the supersymmetry if
$(p,q)$ are co-prime~\cite{Douglas,Polch,ILPT}.
If the 2-brane is wrapped round $S_{rs}$ and the 4-brane is wrapped round
$S_{rs}\times T^2$,
this gives rise to a BPS dyon in four dimensions with charge  $(p,q)$
preserving
$1/4$ of the original Type II 32 supersymmetries.
These all become massless at the same point in moduli space when the area of
$S_{rs}$ tends to zero~\cite{HTE}. Alternatively, one can compactify  first to
8
dimensions on $T^2$, and wrap
a $D=8$
$(p,q)$ dyonic membrane~\cite{ILPT} round $S_{rs}$ to get the same dyonic
state when
reducing further on
$N_4$.

\section{Multiple Branes}
\vskip 3mm
\noindent{\it Intersecting Branes in Flat Space}
\vskip 3mm
There has been much interest in configurations of multiple branes that preserve
some
supersymmetry. A $p$-brane and a $q$-brane can intersect
or overlap 
in  an
$r$-brane and be at an angle
$\theta $ to one another. The amount of   supersymmetry preserved depends on
$p,q,r$ and
$\theta$, and on  the configuration of the D-brane world-volume Yang-Mills
field~\cite{Polch,Green}.  For example,  a 2-brane and a 4-brane of the Type IIA theory
can intersect in a
string.
If they do so perpendicularly ($2\perp 4$, $\theta=\pi/2 $),
the configuration
preserves $1/4$ of the
supersymmetry~\cite{Polch,Green,BDL,Gaunt,Gauntrev}.
For  $\theta=0$, so that the
2-brane lies inside the 4-brane ($2\subset 4$), the (non-threshold)
configuration
preserves $1/2$ of the supersymmetry,
with a magnetic monopole  configuration on the
 4-brane~\cite{Douglas,Polch,BDL}.
The corresponding supergravity solution~\cite{ILPT}
  is
parameterised by an  angle parameter $0\le \zeta\le \pi/2$.
For   $\zeta=0$
the solution is  a single 4-brane, for
$\zeta=\pi/2$ it is a 2-brane, while for general $\zeta$ it is a
``dyon'' carrying both 2-brane and 4-brane charge~\cite{ILPT}.
On reducing on $T^2$, these give the dyonic membranes of~\cite{ILPT}, obtained
by acting on
an electrically charged  membrane with a $U(1)$  subgroup of the $D=8$
U-duality  group, parameterised by the angle $\zeta$.

For $\theta\ne \pi/2,0$ the configuration still preserves $1/4$ of the
supersymmetry,
provided  additional background fields and/or branes are turned on.
The corresponding supergravity solutions, which include $2\perp 4$ and
$2\subset 4$ as
special cases was given in ~\cite{CC}.  The general case  can be interpreted as
a $(2\perp 4|2)$
BPS
(non-threshold) configuration~\footnote{Here the notation $(2\perp 4|2)$
denotes  a
composite configuration of two 2-branes and one 4-brane,
 with $2\perp 4$   refering to the original configuration
 on which  a specific boost and  a subset of U-duality transformations were
 performed in order to arrive at the  final configuration.}.
  It is specified by {  two} independent harmonic
functions $H_{1,2}$ and the angle $\theta=\pi/2-\zeta$.
For  $\theta=\pi/2$ it reduces to $2\perp 4$ ($1/4$ of the supersymmetry)  and
for
$\theta
=0$ it becomes a single 2-brane (It depends on one  single harmonic function
related to $H_{1,2}$.).  When  one of the two original  harmonic functions is
turned
off, i.e.  $H_1=1$, the configuration (preserving $1/2$ of the supersymmetry)
has an  interpretation
as $2\subset 4$  for   all $\theta$. (For  $\theta =0,\pi/2$  it becomes  a
single
2-brane
and a single 4-brane, respectively.)

It will be useful to review the supersymmetry constraints  for the
$2\perp
4$ configuration, which is specified by the harmonic functions $H_{1,2}$ for
the
2-brane and 4-brane, respectively. Suppose that the 2-brane lies in the
$(X^1,X^4)$ plane, and the
4-brane lies in the $(X^2,X^4,X^5,X^6)$ plane of the Minkowski space
$\R^{(9,1)}$.
The  respective supersymmetry constraints  for the 2-brane and 4-brane are
\cite{Polch,Gaunt,Green}:
\begin{eqnarray}
\Gamma^0\Gamma^{1}\Gamma^{4}\epsilon&=&\epsilon,\cr\nonumber
\Gamma^0\Gamma^{2}\Gamma^{4}\Gamma^{5}\Gamma^{6}\ggg^{11}
\epsilon&=&\epsilon.
\label{gammaflat}
\end{eqnarray}
These constraints  (\ref{gammaflat}) are compatible because the constraints
commute,
\begin{equation}
[\Gamma^0\Gamma^{1}\Gamma^{4},\Gamma^0\Gamma^2\Gamma^4\Gamma^5\Gamma^6\ggg^{11}]= 0.
\label{gammacom}
\end{equation}
Thus the configuration $2\perp 4$ is supersymmetric, preserving $1/4$ of
the
supersymmetry.

On the other hand, for the $2\subset 4$ configuration
with the 2-brane lying in the $(X^1,X^2)$ plane, and the
4-brane   in the $(X^1,X^2,X^3,X^4)$ plane, the 2-brane
 and 4-brane supersymmetry constraints
would be:
\begin{eqnarray}
\Gamma^0\Gamma^{1}\Gamma^{2}\epsilon&=&\epsilon,\cr\nonumber
\Gamma^0\Gamma^{1}\Gamma^{2}\Gamma^{3}\Gamma^{4}
\ggg^{11}\epsilon&=&\epsilon.
\label{gammaflata}
\end{eqnarray}
In this case, the constraints anti-commute
\begin{equation}
\{\Gamma^0\Gamma^{1}\Gamma^{2},\Gamma^0\Gamma^{1}\Gamma^{2}\Gamma^{3}\Gamma^{4}\ggg ^{11}
\}= 0,
\label{gammacoma}
\end{equation}
and the combined 2-brane and 4-brane bound state preserves the
supersymmetries with parameters satisfying
\begin{equation}
\left( \cos\zeta \Gamma^0\Gamma^{1}\Gamma^{2}
+\sin \zeta \Gamma^0\Gamma^{1}\Gamma^{2}\Gamma^{3}\Gamma^{4}
\ggg ^{11}
\right)
\epsilon = \epsilon.
\label{gammaflatas}
\end{equation}
There is one constraint and so the configuration
 breaks only half the supersymmetry \cite{ILPT}.

For  a general configuration $(2\perp 4|2)$  with $H_1\ne 1$ and $\theta\ne (0,
\pi/2)$
the supersymmetry constraints are more
complicated,  involving  additional non-zero background fields. Such a
configuration
can be  obtained    by starting with a known BPS configuration, e.g., $2\perp
4$,
and
performing a set of boosts and $U$-duality transformations~\cite{CC}. The
original
configuration  has $1/4$ of the  supersymmetry, and since the  imposed
transformations
are {consistent} with the   supersymmetry transformations, the resulting
configuration continues to
preserve  $1/4$ of  the supersymmetry.

 We now
 extend this   to (multiple) branes wrapped round homology cycles of
some  space. We will focus here on 2-branes and 4-branes  on the space
$N_4\times T^2$ where $N_4$ is an ALE or ALF
instanton,  but the generalisation to other spaces and other branes should be
straightforward.
\vskip 3mm
\noindent{\it  Same Type $p$-branes Wrapping Different Homology Cycles}
\vskip 3mm
Consider
first states arising from two 2-branes wrapped round two
different cycles of $N_4\times T^2$.
One 2-brane is wrapped on $S_{rs}$ and one wrapped on
$S_{tu}$.
Each is located at a particular point in $\R^{(3,1)}\times T^2$ and on
compactification to
$D=4$  Minkowski space $\R^{(3,1)}$
give rise to two  0-branes
 located at two different points in $D=4$.  This gives two
electrically charged  ``W-boson''  vector supermultiplets, one of which is
electrically
charged with respect
to $A^{rs}$ and
 becomes massless as
$S_{rs}$ shrinks and the other   of which is electrically charged with respect
to $A^{tu}$ and
 becomes massless as $S_{tu}$
shrinks. This is a two-particle state, and no bound state is expected, even
though there is no
static force between the two multiplets. The amount of supersymmetry preserved
depends on the
complex structures of the two two-cycles. The one-particle state from a 2-brane
on
$S_{rs}$ preserves half of the  $N=4$ supersymmetries if
$S_{rs}$ is holomorphic, while the   2-brane on
$S_{tu}$  also preserves half of the supersymmetry if
$S_{tu}$ is holomorphic; the combined two-particle state preserves half the
$N=4$ supersymmetries
{ if $S_{rs}$ and
$S_{tu}$ are holomorphic with respect to the {\it same } complex structure}
which,  as  explained in the previous section, requires the vectors $x_r-x_s$
and
$x_t-x_u$ in $\R^3$ to be
parallel.
Choosing the two two-cycles to be holomorphic with aligned    complex
structures
allows a smooth
deformation away from the enhanced symmetry point without any change in the
amount of
supersymmetry preserved.

Similarly, choosing two 4-branes, one wrapped on $S_{rs}\times T^2$ and one
wrapped on
$S_{tu}\times T^2$, gives a two-particle state  with two
magnetic monopoles and
  will preserve half the $N=4$ supersymmetry if both   $S_{rs}$ and
$S_{tu}$ are holomorphic with respect to the   same  complex structure. Again,
as
both  two-cycles shrink to zero, this configuration becomes  massless.

The above  examples of   same-type $p$-branes ($p=2,4$)
 wrapping different cohomology cycles are  supersymmetric configurations
preserving
half  of the supersymmetry (provided the $S_{rs}$ and $S_{tu}$ are holomorphic
with
respect to the same complex structure), but
 {   no bound state is expected}, even
though there is no
static force between the two multiplets.
This is confirmed by the heterotic string picture.
The first set of states
corresponds to perturbative string states of toroidally compactified heterotic
string
with the charge lattice vector length-squared   $-4$, which  are not in the
spectrum of a single string (which requires length-squared not less than
$-2$), but are in the
two-string spectrum. The second set of states are (magnetic)
$\Z_2$  S-duals of the perturbative string states  and are thus also absent in
the single
string spectrum.
\vskip 3mm
\noindent{\it  2-brane and 4-brane  Wrapping Different Homology Cycles}
\vskip 3mm
Consider now the more interesting case in which a 2-brane is wrapped on
$S_{rs}$ and a
4-brane is wrapped on
$S_{tu}\times T^2$,
giving rise to a composite (two-particle) state in general. One particle is
electrically
charged with respect
to $A^{rs}$ and
 becomes massless as
$S_{rs}$ shrinks and the other    is magnetically charged with respect
to $A^{tu}$ and
 becomes massless as $S_{tu}$
shrinks. If $S_{rs}$ is holomorphic, the 2-brane satisfies the supersymmetry
constraints  (\ref{scI}) and (\ref{scIII}), while if $S_{tu}$ is
holomorphic the 4-brane
preserves the
supersymmetry constraints  (\ref{scI}) and (\ref{scIV}).
Let  the
angle between the vectors
$x_r-x_s$ and
$x_t-x_u$ in $\R^3$ be $\theta$.
The analysis of the surviving supersymmetry is similar to that of a 2-brane and
a 4-brane
intersecting in a string at an angle $\theta$. In particular, for
$\theta=\pi/2$ the supersymmetry constraints
(\ref{scIII}) and
(\ref{scIV}) can be satisfied simultaneously, providing the
{  $S_{rs}$ and
$S_{tu}$ are holomorphic with respect to    orthogonal complex structures}
which  requires the vectors $x_r-x_s$ and
$x_t-x_u$ in $\R^3$  to be {  orthogonal}. This is because, in order to satisfy
(\ref{scIII})
and (\ref{scIV}) simultaneously,
the $\Gamma$ matrices have to satisfy the  following
constraint:
\begin{equation}
[\Gamma^0n_{rs}^i\Gamma^i \Gamma^4,
\Gamma^0n_{tu}^j\Gamma^j\Gamma^4\Gamma^5\Gamma^6\Gamma^{11}]= 0,
\label{gammasim}
\end{equation}
which is satisfied if
\begin{equation}
\sum_{i=1}^3 n_{rs}^in_{tu}^i=0.
\end{equation}
For say, $n_{rs}=(1,0,0)$ and $n_{tu}=(0,1,0)$,
the  supersymmetry constraints (\ref{scIII}) and (\ref{scIV}) are completely
analogous to  constraints (\ref{gammaflat}) of a 2-brane and  a 4-brane
intersecting orthogonally
 in a string.

This  analysis can be generalised to give dyonic two-particle states so that a
2-brane
inside a 4-brane wraps
$S_{rs}\times T^2$ to give a $(p,q)$ dyon and another 2-brane inside a 4-brane
wraps  $S_{tu}\times
T^2$ to give a $(p',q')$ dyon. Alternatively one can start from $D=8$
Type II theory on
$\R^{(3,1)}\times N_4$ and wrap a $(p,q)$ membrane on $S_{rs}$
and a $(p',q')$ membrane on $S_{tu}$.
Note that the  supersymmetry constraints  (\ref{scIII}) and (\ref{scIV}), are
modified if one turns on background fields, and in this case it may be possible
to
satisfy these supersymmetry constraints also in the case when  the   vectors
$x_r-x_s$ and
$x_t-x_u$ in $\R^3$  point in directions that are {  not  orthogonal}.

%%%%%%%%%%%%%%

\section{Comparison with Heterotic String Solutions }

The BPS states constructed above  by wrapping branes on
cycles of $N_4\times T^2$,
where
$N_4$ is an ALE space,  have analogues    for branes on $K3\times T^2$.
Type II string theory
compactified on
 $K3\times T^2$ is equivalent to the heterotic string compactified on $T^6$ and
the low-energy
effective field theory is $N=4$ supergravity coupled to 22 vector multiplets.
The BPS 0-brane
states can be associated with supersymmetric spherically symmetric supergravity
solutions.

Perturbative states of the heterotic string carry electric charges ${\vec
\alpha}$ that take
values in an
even
self-dual   lattice  $\Lambda_{6,22}$  and have length-squared (with respect to
the $O(6,22,\Z)$ norm)
 ${\vec \alpha}^TL{\vec\alpha}=-2,0, 2,4,\cdots $. States
 carrying only magnetic charges  ${\vec\beta}$ are related to these
by S-duality, so that
the magnetic charges also take values in the
even
self-dual   lattice  $\Lambda_{6,22}$~\cite{SEN} and have length-squared
 ${\vec \beta}^TL{\vec\beta}=-2,0, 2,4, \cdots$.
States carrying magnetic charges are non-perturbative
and arise from solitons or
black hole
solutions of the theory.
 The black holes in medium-short multiplets which
  become massless at certain points of moduli space have electric
charge vector ${\vec \alpha}$ and magnetic charge vector ${\vec \beta}$
satisfying
${\vec \alpha}^TL{\vec\alpha}=
{\vec \beta}^TL{\vec\beta}=
-2$~\cite{ChC} and the question arises as to whether these are single-particle
quantum states of the non-perturbative heterotic string;
if so, then there are gravitini that become
massless, e.g.,  at the
$SU(2)\times
SU(2)$ or $SU(3)$ enhanced symmetry points~\cite{ChC}  of the $T^2$ moduli
subspace,
so that as well as an enhancement of
gauge symmetry,
there is an enhancement of the supersymmetry.
These medium-short multiplets fit into an infinite $SL(2,\Z)$ representation,
and all the
$SL(2,\Z)$ partners become massless at the same time as the original multiplet.
Each contains a
gravitino, so an infinite number of gravitini would become become massless
simultaneously in this
scenario.
Unfortunately we do not know
the full physical
state conditions on the electric and magnetic charges
for the non-perturbative heterotic theory, which is why
it is useful to compare these results  with the dual Type II picture
 and use insights from the
wrapped brane description of BPS states.

\vskip 3mm
\noindent{\it The  BPS Mass Formula}
\vskip 3mm
 The BPS mass formula, written
in terms of states of heterotic string compactified on $T^6$, is of the form
{}~\cite{CYI,CTII}~\footnote{
We use the notation  and conventions,  as  specified  in  Refs.
\cite{CYI,CYII},
following, {e.g.},   \cite{SEN}.}:
\begin{equation}
M_{BPS}^2 ={\textstyle{1\over 2} }e^{-2\phi_\infty}
{\vec \beta}^T \mu_R{\vec \beta} +
{\textstyle{1\over 2}}e^{2\phi_\infty}{\vec {\tilde\alpha}^T}
 \mu_R{\vec {\tilde\alpha}}
+ \left [({\vec \beta}^T\mu_R{\vec \beta})
({\vec \alpha}^T\mu_R{\vec \alpha})-({\vec \beta}^T\mu_R
{\vec \alpha})^2\right]^{1\over 2},
\label{ME}
\end{equation}
where
\begin{equation}
{\vec {\tilde \alpha}}\equiv{\vec \alpha}+\Psi_{\infty}{\vec\beta}, \ \
\mu_{R,L}\equiv M_{\infty}\pm L.
\end{equation}
The  (quantised) charge vectors $\vec \alpha$ and $\vec \beta$ lie on a even
self-dual (Narain) lattice $\Lambda_{6,22}$~\cite{SEN}.
The subscript
 $\infty$ refers to the asymptotic ($r\to\infty$)
  value of the  corresponding fields.  The
 moduli matrix $M$
and the dilaton-axion field
$S\equiv\Psi+i{\rm e}^{-2\phi}$ transform covariantly
(along with the  charge vectors ${\vec \alpha},{\vec\beta}$) under  T-duality
($O(6,22,\Z)$) and
S-duality ($SL(2,\Z)$), while the BPS mass formula (\ref{ME})
  remains
invariant under these transformations. In the following we shall drop the
$\infty$ subscript.
\vskip 3mm
\noindent{\it  BPS States with $1/2$ of the Supersymmetry}
\vskip 3mm
 When the
magnetic  and electric  charge vectors
are parallel,  $\vec{\beta}\propto \vec{\alpha}$,
the BPS mass formula (\ref{ME})  is that of the
BPS-saturated states which  preserve $1/2$ of $N=4$ supersymmetry
(see, { e.g.},~\cite{SEN}). These states fit into  ultra-short multiplets,
whose
highest
spin  is one.

The  electric   charge vectors
 have length-squared
$ {\vec \alpha}^TL{\vec\alpha}=-2,0, 2,4, \cdots$ with respect to the
  $O(6,22)$-invariant metric $L$.
States carrying only magnetic charges are related to  the electric states by
S-duality   and the charge quantisation condition then
implies~\cite{SEN}
that their  magnetic charge vectors also
lie on an even self-dual
   lattice  $\Lambda_{6,22}$  with  norm
$
{\vec \beta}^TL{\vec\beta}=-2,0, 2,4,\cdots\ .
$
In addition there is an infinite number of  dyonic states, preserving $1/2$ of
the
supersymmetry, which are  related  to the purely electric states by
S-duality.
Their  charge vectors are proportional,    ${\vec \alpha}\propto
{\vec\beta}$,  with     relatively co-prime electric and magnetic charges.
Purely electrically charged states are in the perturbative string spectrum,
while
dyons and  purely magnetically charged states are  non-perturbative BPS
configurations.

 As classical supersymmetric solutions these states are spherically
symmetric configurations.  (i)  Solutions with    positive electric
length-squared
${\vec \alpha}^TL{\vec\alpha}>0 $  (and its accompanying $SL(2,\Z)$ tower) have
 null
singularities, i.e. the horizon and the  curvature
singularity coincide, so that the horizon is singular. (ii) Solutions
with ${\vec \alpha}^TL{\vec\alpha}=0 $ (along with its   $SL(2,\Z)$ tower)
have
naked singularities, so that a null  probe reaches the singularity
 in  a finite time as measured by an asymptotic observer, while
the gravitational potential for massive test particles is attractive. For
example,
the well known KK
monopole solution is in this class.
These  two classes of solutions  have   mass (\ref{ME}) which is always
positive.
 (iii) Solutions with ${\vec \alpha}^TL{\vec\alpha}=-2 $  (and  the
corresponding $SL(2,\Z)$ tower)   have typical naked
singularities, i.e. a null  probe  again reaches the singularity
 in  a finite  time as measured by an asymptotic observer, but now
the gravitational potential for massive test particles is  repulsive~\footnote{
Qualitative features  of these space-times are  similar to  those of
the  Schwarzschild  black hole solutions with   ``negative mass'' ($M<0$).
These
solutions  have naked singularities that repel massive test particles.}.
 They become massless at
 points of moduli space
where ${\vec\alpha}^T\mu_R{\vec\alpha}=0$.
As these fit into vector multiplets, when they become massless, there
is an enhancement of the
gauge symmetry~\cite{HTE}.
For example, at special points  on the $T^2$ moduli subspace the  gauge  group
enhancements are  of
the type
$U(1)^4\to U(1)^3\times SU(2)$,
$U(1)^4\to  U(1)^2\times SU(2)\times SU(2)$ or $ U(1)^4\to U(1)^2\times SU(3)$.

On the Type II string  side the appearance of such massless states  is due to
wrapping of
$p$-branes round   shrinking two-cycles~\cite{HUSC,HTE,BSV,strom}. They can
carry either an
electric charge $q$, or a magnetic charge $p$, or can be dyonic with charge
$(p,q)$.
 These  dyonic BPS states arise from a 2-brane with charge $q$  wrapped round
a
two-cycle $S$ of
$K3$ and a 4-brane with charge $p$ wrapped round $S\times T^2$
and, as they preserve half the $N=4$ supersymmetry, they fit into ultra-short
vector multiplets. In the quantum theory, the charges
 $(p,q)$ are
co-prime    integers so that the states fill out an $SL(2,\Z)$ representation.
The purely electric
states are the ``W-bosons'' that become massless when the two-cycle $S$
degenerates
and  one of the $U(1)$ gauge symmetries is enhanced to $SU(2)$;
at the same point
in moduli space the infinite set of dyonic partners also become massless.
\vskip 3mm
\noindent{\it BPS States Preserving $1/4$ of the  Supersymmetry}
\vskip 3mm
In the case when the magnetic and
electric charge vectors  are not parallel, the  the BPS mass  is larger than
that of
BPS states preserving $1/2$ of the supersymmetry (the last term in
(\ref{ME}) is non-zero).  These states  are always {  non-perturbative}
and
preserve only $1/4$ of the  $N=4$ supersymmetry~\cite{CYI}. This means that
the corresponding quantum states should fit into medium-short  multiplets of
$N=4$
supersymmetry which include states of at least spin $3/2$; it is expected that
the multiplet
includes scalars so that $3/2$ is in fact the maximum spin.

The classical solutions  correspond to the following types of spherically
symmetric
solutions. (i) Solutions
with both charge norms positive, i.e.
${\vec \alpha}^TL{\vec\alpha}>0$, ${\vec \beta}^TL{\vec\beta}>0$,
 have regular horizons with {\it non-zero, moduli independent,
Bekenstein-Hawking
entropy}~\cite{CTII}:
$\pi[({ \vec \alpha}^TL{\vec\alpha})({\vec \beta}^TL{\vec\beta})-({\vec
\alpha}^TL{\vec\beta})^2]^{1/2}.$
(ii) Hybrid solutions  with either  electric or magnetic charge norm zero
[negative]
have null singularities [naked singularities].
(iii) Solutions with both electric and magnetic charge norms  negative, i.e.
${\vec \alpha}^TL{\vec\alpha}={\vec \beta}^TL{\vec\beta}=-2$, have again
naked singularities, and in addition  can become massless at special points of
moduli space for which ${\vec\alpha}^T\mu_R{\vec\alpha}=0,\ \
{\vec\beta}^T\mu_R{\vec\beta}=0,$ and
${\vec\beta}^T\mu_R{\vec\alpha}=0$.

The latter set of solutions is   of special interest. If there are
corresponding
 quantum states with the same charge vectors, they must
 preserve
$1/4$
of the supersymmetry
and so fit into spin-3/2 super-multiplets.
Moreover, they must become massless
  at special points of moduli space,
giving an infinite number of massless spin-3/2 quantum states and the
possibility of supersymmetry
enhancement~\cite{CYII}.

For example,   on a
  $T^2$ subspace, the  moduli
matrix $M$ and $O(2,2)$ invariant matrix $L$ are  of the form:
\begin{equation}
M=\left ( \matrix{G^{-1} & -G^{-1}B \cr
-B^T G^{-1} & G + B^T G^{-1}B}
\right ), \ \ \  L =\left ( \matrix{0 & I_2\cr
I_2 & 0} \right ),
\end{equation}
 where $G \equiv [{G}_{mn}]$ ($(m,n)=1,2$), $B\equiv [B_{12}]$
 are the four moduli on
$T^2$. The BPS mass formula (\ref{ME}) implies
that these spin-$3/2$ multiplets  become massless  at points where
$U(1)^4$
is enhanced
to $U(1)^2\times SU(2)\times
SU(2)$ or $U(1)^2\times SU(3)$~\cite{ChC}.
These two symmetry enhanced points  take place at the
point of moduli $(G_{11},G_{22},G_{12},B)=(1,1,0,0)$ (the self-dual point of
the two-circle) and
 $(G_{11},G_{22},G_{12},B)=(1,1,{1 \over 2},{1 \over 2})$, respectively.
 The massless dyonic  solutions
at these points of moduli space have the following  respective 
charge vectors~\cite{ChC}:
\begin{equation}
 \ \ \ \ \ \ \ \ \ \ \ \ ({\vec \alpha},{\vec \beta})=({\vec\lambda}_{1\,\pm},
{\vec\lambda}_{2\,\pm}), \ \
\label{dy1}\end{equation}
and
\begin{equation}
\ \ \ \ \ \ \ \
({\vec \alpha},{\vec \beta})=({\vec\lambda}_{i\,\pm},{\vec\lambda}_{j\,\pm}),
\ \  [(i,j)=2,3,4,\ \ i< j],
\label{dy2}
\end{equation}
where:
\begin{equation}
{\vec {\lambda}}_{1\,\pm}\equiv\pm (1,0;-1,0),
{\vec {\lambda}}_{2\,\pm}\equiv\pm (0,1;0,-1),
{\vec{\lambda}}_{3\,\pm}\equiv\pm (1,1;-1,0),
{\vec{\lambda}}_{4\,\pm}\equiv\pm (1,0;-1,1).
\label{lambdai}
\end{equation}

\vskip 3mm
\noindent{\it Correspondence between  the Dual  Descriptions}
\vskip 3mm
The   configurations described above  arise from  the Type II string on
$K3\times T^2$ as
follows.
Let $S$ and $S'$ be supersymmetric two-cycles of $K3$. Wrapping a 2-brane on
$S$
and a 4-brane on
$S'\times T^2$ gives a  $D=4$ two-centre solution  with one   carrying
the electric charge
associated with 2-brane  wrapping  on $S$ and the other carrying the magnetic
charge
associated with 4-brane wrapping on
$S'$.
These will preserve $1/4$ of the heterotic supersymmetry if the complex
structures on $S,S'$
are correlated appropriately, i.e the condition reduces to
the orthogonality condition (\ref{gammasim})
in the  limit of shrinking two-cycles of $K3$ in which the $K3$ can
 be approximated by an ALE space.

 When the two centres coalesce (in $D=4$),  the dual description of the
particular BPS solution emerges.
E.g., choosing a particular hyper-surface  of the $T^2$:
$(G_{11},G_{22},G_{12},B)=(G_{11},G_{22},0,0)$, the BPS
configurations  with  the dyonic charge vectors (\ref{dy1})  have the BPS
mass~\cite{CYII}:
\begin{equation}
M_{BPS}= e^\phi|G_{11}^{1/2}-G_{11}^{-1/2}| +
 e^{-\phi}|G_{22}^{1/2}-G_{22}^{-1/2}|.
\label{massorth}\end{equation}
Along this hyper-surface the configurations are massive,  but become
massless at
$G_{11}=G_{22}=1$  (the $SU(2)\times SU(2)$ point).

The  Type II construction shows that
 the single-centre solution are
in fact the
coincident limit of  two-centre solutions of the $D=4$ heterotic
supergravity theory.
On the  Type II string side we have a
 2-brane and a 4-brane    wrapping round    different  shrinking
 two-cycles,  whose complex structures are orthogonal. The  mass is
proportional
to the sum of the  area of the two (shrinking) two-cycles.  The comparison with
the
dual solution on heterotic string side  above (i.e.  along the specific
hyper-surface
$(G_{11},G_{22},0,0)$  with the BPS mass (\ref{massorth})), yields  the
following
correspondence:  the area of the  two shrinking two-cycles is proportional to
 $|G_{11}^{1/2}-G_{11}^{-1/2}|$ and  $
  |G_{22}^{1/2}-G_{22}^{-1/2}|$,  respectively and the fact that the complex
structures of the two
  two-spheres are orthogonal corresponds on the heterotic  string side to  the
moduli
choice
 $G_{12}=B=0$.

At the  $SU(3)$ point,  on the heterotic string side we chose a
particular  hyper-surface
with the moduli $(G_{11},G_{22},G_{12},B)
 =(G,G,G/2,G/2)$ . The twelve BPS states with dyonic charge vectors (\ref{dy2})
  all
  have the  {\it same}
  BPS mass:
\begin{equation}
M_{BPS}=\left[\textstyle{2\over
3}(e^{2\phi}+e^{-2\phi}+\sqrt{3})\right]^{1/2}|G^{1/2}-G^{-1/2}|.
\label{masssu3}
\end{equation}
For   $G=1$ (the $SU(3)$ point), these configurations  have zero mass.

On the Type II string side  these twelve configurations  correspond to
2-brane and 4-brane pairs, each wrapping round a different
two-cycle, and the three two-cycles   have complex structure directions
that  form an equilateral triangle. The area  of  each of the three two-cycles
is
thus the {\it same}; it  in one to one correspondence with  the value of
$|G^{1/2}-G^{-1/2}|$ on the heterotic string side.
In the heterotic description, there is  a  two-form field on $T^2$ turned on,
i.e. $B=G_{12}=G/2$,
so that there should be a non-trivial two-form gauge field on the
Type II   side also.

\section{Discussion: Two-Particle BPS States and Supersymmetry Enhancement}

 We have seen that the $N=4$ supergravity theory arising from
$D=4$ toroidally compactified heterotic string
  has
(singular) black hole solutions
 that
 (i) carry electric charge with respect to one vector field and
a magnetic charge
  with respect to another and (ii) preserve only $1/4$ of the $N=4$
supersymmetry. This means that
the corresponding quantum states should fit into medium-short  multiplets of
$N=4$
supersymmetry which include states of at least spin $3/2$; it is expected that
the multiplet
includes scalars and $3/2$ is in fact the maximum spin.
Moreover,  the BPS mass formula implies that these multiplets
  become massless when the
$U(1)\times U(1)$ with respect to which the   solution is charged gets enhanced
to $SU(2)\times
SU(2)$ or $SU(3)$.
This led   to the conjecture  that the supersymmetry could be enhanced at these
points in
moduli space, due to
the spin-$3/2$ states in these BPS multiplets becoming massless~\cite{CYII}. In
fact, there would be an
infinite number of extra massless gravitini corresponding to the possible
values of the electric
and magnetic charges.
 The  key question   is whether these are single-particle (quantum)
states of the non-perturbative heterotic string; unfortunately we do not know
the full physical state conditions on the electric and magnetic charge vectors
for the non-perturbative theory~\footnote{In~\cite{DVV} a formula for the
degeneracy of states of $D=4$,  $N=4$  string theory is presented. It is
unfortunately
not suitable for  states with the norm of the charge vectors negative.}.

These $D=4$ solutions are limiting cases of two-centre solutions in which the
centres coincide.
This is particularly clear from the
Type II perspective, in which they arise from two different branes on two
different two-cycles, and
which in general will give two  0-branes at different points
in $D=4$ space-time.
If the corresponding two BPS states form a bound state, this will fit into a
spin-$3/2$ multiplet
that becomes massless, leading to enhanced supersymmetry.
If they do not form a bound state, then there is no enhanced supersymmetry. A
two-particle state constructed from two vector particles can have spin greater
than one, but
when they both are massless, there is no new higher-spin gauge symmetry.

The   structure of the two-particle Type II spectrum     is
perhaps clearest from
the
$D=8$
perspective, i.e. reducing from $D=10$ (Type II)  or $D=11$ (M theory) on $T^2$
or
$T^3$,
respectively.
A dyonic membrane in $D=8$ is obtained from a D-2-brane inside a D-4-brane
 wrapped on $T^2$,
or from an M-2-brane inside a M-4-brane wrapped on $T^3$.
We compactify the $D=8$ Type II theory on $K3$  and choose the example
with a $(p,q)$ membrane
wrapping $S$ and a
$(p',q')$ membrane wrapping $S'$.
The amount
of supersymmetry depends on the ``alignment'' of the  complex structures
$J,J'$ of $S,S'$ (which
are both supersymmetric and so holomorphic with respect to $J,J'$,
respectively)
and of the values
of $(p,q,p',q')$.
The
$\Lambda_{2,2}$ charge lattice is acted on by
  $SL(2,\Z)
\times SL(2,\Z)$, but only the diagonal $SL(2,\Z)$ is a symmetry of the theory.
 It will be
convenient to use
this $SL(2,\Z)$ duality  to set the magnetic charge of the second membrane   to
zero,
$p'=0$.
If $p$ is also zero so that the solution is purely electric, then the
configuration will
break half the $N=4$ supersymmetry if $J,J'$ are aligned, so that they are
 parallel in the
ALE limit.
This is related by $SL(2,\Z)$ duality to a purely
magnetic configuration with
$q=q'=0 $ arising from two D-4-branes, one wrapping
$S\times T^2$
and one wrapping $S '\times T^2 $
(together with an
infinite number of dyonic partners).

If instead $p'=0$ and $q=0$, the configuration will
break $1/4$ of the heterotic supersymmetry if $J,J'$ are arranged so that they
are  orthogonal in the
ALE limit.
This is the set-up that gives, in the coincident limit,  the spherically
symmetric
solution   (with naked singularity), fitting into a
medium-short multiplet. Acting with $SL(2,\Z)$ duality then gives an infinite
number of
partners to this, all of
which become massless
(with a corresponding choice of the NS-NS two-form gauge field expectation
value
on $K3$~\cite{Aspin,Aspinwall}).
As $S$ and $S'$   shrink, the limit gives $SU(2)\times SU(2)$
enhanced gauge
symmetry. If  the two-cycles    intersect as they
degenerate, an $SU(3)$  enhanced
symmetry results ($A_2$ singularity)~\cite{BSV}.  In
each case there  are two-particle  BPS states.

For the purely electric case, the state with charges $q=q'=1$,
$p=p'=0$ represents two ``W-bosons'', one for the $SU(2)$ associated with $S$
and
one for
that associated with $S'$, and  there is no bound state. Indeed such a bound
state
would be problematic for the   Higgs mechanism
picture of  gauge symmetry enhancement. As these states are electrically
charged,
they
can be analysed
from the point of view of the perturbative heterotic string.
The  length-squared of these charge vectors is $-4$, so that  they are  not a
physical
states of a single
 heterotic string, but do  occur in the two-string spectrum.

Consider now the case in which $p'=0$ and $q=0$, together with its S-dual
partners.
If there is a bound state, then there is an infinite enhancement of
supersymmetry when $S,S'$
degenerate.
Since there are an infinite number of  dyonic states (S-dual to the
``W-bosons'')
that
become massless  at the points of enhanced gauge symmetry, this limit cannot be
described by a field theory. Thus the infinite  number of  massless states
which
preserve $1/4$ of the supersymmetry,  which would include
 the infinite number of massless  gravitini with dyonic charges, cannot be
immediately ruled out.
The massless states  should then fit into representations of some
supersymmetry algebra, and if it were some conventional $N>4$ algebra, it
seems
likely that
this would require an infinite massless multiplet with arbitrarily high spin.
This would be  worse than the decompactification limit of a Kaluza-Klein
theory, which also has
an infinite number of particles  becoming massless, but  with the maximum spin
remaining two.
This kind of spectrum might   result from the tensionless (null) limit of
something like a
non-critical
superstring theory~\cite{Hanany}.
Note however, that
such dramatic behaviour occurs at, e.g.,  an
$SU(2)\times SU(2)$
enhanced symmetry point, but not at an $SU(2)$ one.

However, it is also possible that the supersymmetry enhancement could be
associated with a
decompactification limit of the type discussed in~\cite{Scherk}. If a string
theory in
$D+1$
dimensions is
compactified on a circle, masses can be introduced via a Scherk-Schwarz
mechanism in such a way that
supersymmetry is spontaneously broken~\cite{Scherk}. In string theory, the
limit in
which
the circle shrinks
to zero size is a decompactification to a T-dual string theory in $D+1$
dimensions, in which the
supersymmetry can be restored~\cite{KKPR,GKKOPP}. From the point of view 
of the $D$
dimensional
theory, there are
two points at the boundaries of moduli space $(R=0$ and $R=\infty$) at which
(i) infinite towers
of Kaluza-Klein modes become massless and (ii) in addition, extra gravitino
multiplets become
massless, giving extra supersymmetry.
It is possible that there could be  a somewhat analogous supersymmetry
enhancement in the cases
considered here.
 Evidence for the appearance of an infinite number massless
gravitini  at special points in moduli space was  also found
 in the study of BPS states of non-critical strings
which  appear due to the zero-size instantons of exceptional
groups~\cite{KMV}.

If there is no bound state a much simpler if less dramatic picture emerges.
The ``double wrapping states'' are all two-particle states and there is no
enhancement of supersymmetry,
only the by-now familiar picture of  gauge symmetry enhancement. It would be
interesting to check this
directly by studying the quantum mechanics of the two-particle system.
However, the fact that this would give a simple consistent picture whereas the
alternative
requires exotic new physics might seem to make the two-particle picture the
most likely
interpretation of these results.

\acknowledgments
We would like to thank Gary Gibbons for helpful discussions and  Jerome
Gauntlett for pointing out errors in the earlier version of the manuscript.
The work was supported in part by U.S. Department of Energy Grant No.
DE-FG02-95ER40893 (M.C.),  the National Science  Foundation Career
Advancement Award  PHY95-12732 (M.C.) and the NATO collaborative  research
grant
CGR 949870 (M.C.). We would like to thank the Newton
Institute in Cambridge, the
Department of Physics and Astronomy, University of Pennsylvania (C.M.H.)
and the Aspen Center for Physics (M.C.),
for hospitality.

\end{document}